\documentstyle[12pt]{article}

\newcommand{\beq}{\begin{eqnarray}}
\newcommand{\eeq}{\end{eqnarray}}
\begin{document}

\vspace*{1cm}
\begin{center}
{\large Quark mean field model for nuclear matter and finite nuclei\\}
\vspace*{1.5cm}
{H. Shen\footnote{e-mail address: shen@rcnp.osaka-u.ac.jp} \\
Department of Physics, Nankai University, Tianjin 300071, China\\}
\vspace*{0.5cm}
{H. Toki\footnote{e-mail address: toki@rcnp.osaka-u.ac.jp} \\
Research Center for Nuclear Physics (RCNP), Osaka University, Ibaraki,\\
Osaka 567-0047, Japan\\}
\vspace*{0.5cm}
\end{center}
\vspace*{0.5cm}
\abstract{We study nuclear matter and finite nuclei in terms of 
the quark mean field (QMF) model, in which
we describe the nucleon using the constituent quark model.
The meson mean fields, in particular the $\sigma$ meson,
created by other nucleons act on quarks
inside a nucleon and change the nucleon properties in nuclear medium.
The QMF model predicts an increasing size of the nucleon 
as well as a reduction of the nucleon mass in the nuclear environment.
The present model is applied to study the properties of finite nuclei
after fixing all the parameters by the nuclear matter properties,
and it is found to give satisfactory results on the nuclear properties.\\

\vspace*{0.5cm}
\newpage


\section{Introduction}

We have been describing the nucleus with the assumption that
the nucleon properties are unchanged from those of the free space nucleon.
However, after the EMC effect was reported, we encounter often the
discussions on the change of hadron properties in nuclei~\cite{EMC}.
The EMC effect motivated many theoretical works on the study of hadrons
in terms of quarks and gluons, although the pion interpretation
is not excluded~\cite{ericson}.  Among them, the QCD sum rule suggested
a large change of the masses of the vector mesons and nucleons in
nuclei~\cite{hatsuda}.
It is very important to pursue these possibilities both
theoretically and experimentally from various view points.

On the other hand, recent theoretical studies show that
the properties of nuclear matter can be described nicely
in terms of the relativistic Brueckner-Hartree-Fock (RBHF) approach~\cite{brockmann}.
With the nucleon-nucleon interaction fixed from the nucleon-nucleon scattering and the
deuteron properties, we can reproduce very closely the nuclear
matter saturation properties.  The reason of the success is the large
reduction of the nucleon effective mass in nuclear matter, which
provides a density dependent repulsive contribution for
the energy density.  This reduction of the nucleon mass is
also the source of the large spin-orbit splitting in finite
nuclei, which is the key phenomenon of the nuclear shell model.
Any model which changes the hadron properties in nuclear matter should
respect these observations.

Guichon proposed an interesting model
on the change of the nucleon properties in nuclear
matter~\cite{guichon}. The model construction mimics the relativistic 
mean field theory, where the
scalar and the vector meson fields couple not with nucleons but directly
with  the quarks in nucleons of the nuclei~\cite{walecka}.  Hence, the
nucleon properties change according to
the strengths of the mean fields acting on the quarks.  The
nucleon is modeled in terms of the MIT bag model~\cite{MIT}.
The scalar meson provides a strong attraction and as a consequence
provides a negative mass to the quarks.
The MIT bag model then reduces the
nucleon mass in nuclear matter.  This model was extended by Thomas
and his collaborators under the name of the quark-meson coupling (QMC)
model and applied to many observables~\cite{QMC}.

The model of Guichon relies strongly
on the choice of the nucleon model, for which the QMC model
uses the MIT bag approach~\cite{MIT}.  The MIT bag model assumes that
the inside of the bag is the perturbative vacuum and
the quark mass is the bare mass, which is nearly zero for the
up and down quarks.  The perturbative vacuum has a larger energy
than the non-perturbative vacuum and therefore the bag constant
takes care of the energy difference between the two vacua.
The non-perturbative objects as mesons are not allowed to exist
in the bag interior and stay outside the bag.   The chiral
symmetry, the continuity of the axial current, requires the coupling of
quarks with pions at the bag surface~\cite{hosaka}.

There are several conceptual problems to take the MIT bag model in the
Guichon model;\\
1. The non-perturbative objects, $\sigma$ and $\omega$ mesons, have
to be present in the perturbative vacuum.\\
2. The quark mass has to change from its bare mass
due to the coupling to the $\sigma$ meson.\\
3. For the up and down quarks, the resulting mass of the quarks is
negative.\\
\noindent
There may be some arguments to overcome the above mentioned points
and justify the QMC model~\cite{QMC}.

We would rather like to take another model for the nucleon, which
is the constituent quark model~\cite{IK}.
In this model, the quarks get constituent quark masses
due to spontaneous chiral symmetry breaking.  It is then
natural to have nearly zero mass pions as the Nambu-Goldstone bosons.
Their coupling to the constituent quarks is provided by the
chiral symmetry.  This consideration makes a simple interpretation of
the direct coupling of not only pions but also other mesons as
$\sigma$ and $\omega$ mesons, since there is no boundary
to separate the perturbative and the non-perturbative
regions in this picture.  The $\sigma$ mean field could be considered
as the amount related with the change of the chiral condensate
in nuclear medium.  Hence, it is natural to get the reduction
of the quark mass in the nucleon inside of nuclei from
the quark mass of the nucleon in the free space.

The constituent quark model is used extensively also for
nucleon-nucleon interaction and later extended to
baryon-baryon interactions in the SU(3) sector
with great success~\cite{QCM,oka,fae,fuji,str,fae1}.
In this picture, it is
natural that the mesons couple with quarks, since
the nucleon is the composite of quarks.

Hence, it is very interesting to construct the Guichon
model, where the nucleon is described in terms of
constituent quarks, which couple with mesons and gluons.
This model (we refer it as the quark mean field (QMF) model 
as named in Ref.~\cite{QMF}.) has
a direct connection to the one-boson exchange model (Bonn potential).
We expect, therefore, the quantitative results similar to those of the
RBHF theory~\cite{brockmann}.

The paper is arranged as follows.  In Section 2 we describe the concept
of the QMF model and the procedure to perform calculations.
In Section 3 we show the results for nuclear matter and
in Section 4 the results for finite nuclei are presented.
Section 5 is devoted to the summary of this paper.

\section{Quark mean field model}

We are still far away from describing nucleons and nuclei in terms of
quarks and gluons using QCD.  This difficulty arises from
the fact that the theory becomes highly non-perturbative
at low energy ($p < 1GeV$).  In addition, the QCD vacuum
is realized in a non-trivial way, where the chiral symmetry is
broken and the quarks and gluons are confined.  The lattice
QCD (LQCD) may handle hadrons.  It is, however, too much to ask
the LQCD to describe a system of many nucleons, the nuclei.
Hence, we resort to an effective theory of QCD at
low energy, which is based on QCD.  The dual Ginzburg-Landau (DGL) theory 
may be a good candidate~\cite{DGL}.

In this paper, we take a more phenomenological view point.
We shall begin with a possible Lagrangian of the quark many body
system.   For this Lagrangian, we take into account
the consequence of the non-perturbative gluon dynamics of spontaneous
chiral symmetry breaking and quark confinement~\cite{ChenLi}.
The effective Lagrangian of this level may be written as
\beq
 {\cal L} =\bar q \left( i \gamma_\mu \partial^\mu - m_q - \chi_c
 - g \gamma_\mu A^\mu - g_i^q \phi_i \Gamma_i \right) q
 + {\cal L}(\chi_c, A_\mu, \phi_i) \,  .
\eeq
Here, $q$ denotes the quark fields with constituent quark mass $m_q$,
which is of the order of $300 MeV$ to be consistent with
the quark condensate and the pion decay constant~\cite{weise}.
The confinement is expressed in terms of
$\chi_c$, which is given by the gluon dynamics \cite{DGL}.
The path from QCD to this expression is not yet known.
One method might be the use of the dual
Ginzburg-Landau (DGL) theory, which contains QCD monopoles in
the Abelian space in the Abelian gauge
and their condensation to induce the
dual Meissner effect~\cite{DGL}.  $A^\mu$  are the gluon
fields with the running coupling constant $g$ at the model 
scale. From the lattice QCD data, 
we find the effect of the gluon exchange 
interaction below $r \leq 0.2fm$ and hence the running coupling
constant should be taken at the momentum scale, $\mu \sim 1GeV$.
We include meson fields $\phi_i$, which couple with quark fields
with the Dirac matrices $\Gamma_i$ and the
coupling strength $g_i^q$.  The rest ${\cal L}(\chi_c, A_\mu, \phi_i)$
denotes the confinement field, gluon and meson dynamics,
which is not specified here explicitly.
The explicit form is quite involved.

To proceed to many body system,
we take the mean field approximation for the meson
fields. We restrict to $\sigma$, $\omega$, and $\rho$ mesons,
which are commonly used in the relativistic mean field
model~\cite{walecka,hirata,ring}.
We note that in the mean field approximation, the pion field does not
survive due to the spin average.
Since quarks are confined completely in hadrons, we work out the
many body problem in two steps.
First, we construct the nucleon under the
influence of the meson mean fields.
Then in the second step, we solve the entire nuclear system with
the change of the nucleon properties due to the presence of the mean fields,
which is obtained in the first step.

The first step is to generate the nucleon system under the
influence of the meson mean fields.
In the constituent quark model, the quarks in a nucleon satisfies
the following Dirac equation:
\beq
\left[i\gamma_{\mu}\partial^{\mu}-m_q-\chi_c
-g^q_{\sigma}\sigma(r)
-g^q_{\omega}\omega(r)\gamma^0
-g^q_{\rho}\rho(r)\tau_3\gamma^0
\right] q(r) = 0,
\eeq
where $\tau_3$ is the isospin matrix in our nuclear physics convention.
Assuming the meson mean fields are constant within the small nucleon
volume, we can then write the Dirac equation as
\beq
 \left[ -i \vec \alpha \cdot \vec\nabla + \beta m_q^{*} +
 \beta \chi_c \right] q(r)=e^{*} q(r),
\eeq
where $m_q^{*}=m_q+g^q_{\sigma}\sigma$ and
$e^{*}=e-g^q_{\omega}\omega-g^q_{\rho}\rho\tau_3$,
with $\sigma$, $\omega$, and $\rho$ being the mean fields
at the middle of the nucleon.
$e$ is the energy of the quark under the influence of the $\sigma$,
$\omega$,
and $\rho$ mean fields.
The quark mass is modified to $m_q^{*}$ due to the presence of the $\sigma$
mean field. Here, $g^q_{\sigma}$, $g^q_{\omega}$, and $g^q_{\rho}$ are
the coupling constants of the $\sigma$, $\omega$, and $\rho$ mesons
with quarks, respectively. We take into account the spin correlations,
$E_{spin}$, due to gluons and pions so that the mass difference between
$\Delta$ and nucleon arises.
Hence, the nucleon energy is expressed as $E_n^{*}=3e^{*}+E_{spin}$,
where the vector contribution is removed here.
There exists the spurious center of mass motion, which is removed
in the standard method by $M_n^{*}=\sqrt{{E_n^{*}}^2-<p^2_{cm}>}$,
where $<p^2_{cm}>=\sum_{i=1}^{3}<p^2_i>$, since the three constituent quarks
are moving in the confining potential independently.

We now move to the second step, in which the nuclear many body system
will be solved with the change of the nucleon properties obtained in
the first step.  We assume the following QMF Lagrangian,
\beq
 {\cal L}_{QMF}=\bar \psi \left[ i \gamma _ \mu \partial ^\mu -
M_n^* - g_\omega \omega \gamma^0
-g_\rho \rho \tau_3\gamma^0\right]
\psi + {\cal L}_M(\sigma, \omega, \rho)  .
\eeq
Here, $\psi$ denotes the nucleon fields. The change of the nucleon
properties is worked out in the first step and the outcome is exclusively
expressed in $M_n^*$, which is a function of the quark mass
correction, $\delta m_q$, related to the $\sigma$ mean field
as $\delta m_q=-g_{\sigma}^q\sigma$.
The $\omega$ and $\rho$
mean fields do not cause any change of the nucleon properties~\cite{fleck},
and they appear merely as the energy shift. These contributions are carried
over as the nucleon-meson coupling terms with the replacement of the
quark-meson couplings as  $g_{\omega}=3g_\omega^q$ and $g_\rho=g_\rho^q$.
We will apply the QMF model to study the properties of nuclear matter
and finite nuclei in the following two Sections.

\section{properties of nuclear matter}

We calculate first the change of the nucleon properties
as a function of the quark mass correction, $\delta m_q$,
which is defined as
$\delta m_q=m_q-m_q^{*}=-g_{\sigma}^q\sigma$.
Here, the constituent quark mass is taken to be one third of
the nucleon mass; $m_q=M_n/3=313 MeV$.
We take into account confinement in terms of the
harmonic oscillator potential together with two Lorentz structures;
(1) scalar potential        $\chi_c=\frac{1}{2}kr^2$
(2) scalar-vector potential $\chi_c=\frac{1}{2}kr^2 (1+\gamma^0)/2$.
As pointed out in Ref.~\cite{Conf}, the quark can not be confined when
the vector potential is larger than the scalar one.
Here, we just take two extreme types, since the Lorentz structure of
the confinement is not established.
As for the strength of the confining potential,
we take $k=700$ and $1000 MeV/fm^2$, in order to see the results
depending on this factor.
The spin correlation, $E_{spin}$, is fixed by the free nucleon mass as
$M_n=\sqrt{(3e+E_{spin})^2-<p^2_{cm}>}=939 MeV$.
We assume further that the confining interaction
and the spin correlations do not change in the nuclear medium.

We show in Fig.1 the results of the effective nucleon mass $M_n^*$
as a function of $\delta m_q$. 
The results of the scalar potential with two oscillator parameters, 
$k=700$ and $1000 MeV/fm^2$, are shown by the two solid curves.
The effective nucleon mass decreases with $\delta m_q$
and the curvature is clearly negative, which influences
the nuclear matter properties, as will be discussed soon.
The results of the scalar-vector potential with the same two
oscillator parameters are shown by the dashed curves.
In this case the dependence of $M_n^*$ on $\delta m_q$
is almost linear as has been reported by Toki {\it et al}.~\cite{QMF}.
We note that the results of the
nuclear matter are very sensitive to the behavior of $M_n^*$.

To perform the nuclear matter calculation, we use the relativistic
mean field (RMF) approximation containing nucleons,
neutral scalar ($\sigma$)
and vector ($\omega$) and isovector ($\rho$) mesons.
We need to specify the meson Lagrangian ${\cal L}_M $, since
there are various versions of the meson Lagrangian in the RMF theory.
The RMF theory with the TM1 parameter set includes the nonlinear
terms both for $\sigma$ and $\omega$ mesons,
which can reproduce the feature of the RBHF theory
and satisfactory properties of finite nuclei~\cite{TM1}.
In the present model,
we would like to take the meson Lagrangian as
\begin{eqnarray}
 {\cal L}_{M}&= &\frac{1}{2}(\partial_\mu\sigma)^2
                -\frac{1}{2}m_\sigma^2\sigma^2
                -\frac{1}{4}g_3\sigma^4             \\ \nonumber
             & &-\frac{1}{4}(\partial_\mu\omega_\nu
                           -\partial_\nu\omega_\mu)^2
                +\frac{1}{2}m_\omega^2\omega_\mu^2
                +\frac{1}{4}c_3\omega_\mu^4   \\ \nonumber
             & &-\frac{1}{4}(\partial_\mu\rho^a_\nu
                            -\partial_\nu\rho^a_\mu)^2
                +\frac{1}{2}m_\rho^2(\rho^a_\mu)^2.
\end{eqnarray}
Comparing with the Lagrangian in the RMF(TM1) model~\cite{TM1},
we have deleted the nonlinear term
$\frac{1}{3}g_2\sigma^3$, so that less free parameters are contained
in the present model. Actually, the nonlinear term
$\frac{1}{3}g_2\sigma^3$ plays similar role as the nonlinear
term $\frac{1}{4}g_3\sigma^4$.
Note that we have dropped the 
$\rho^a_\mu \times \rho^b_\nu$ term, since this term vanishes at the mean field level.
Now, the nucleons and mesons obey the Euler-Lagrange equations
derived from the QMF Lagrangian, which can
be written, for the nuclear matter, as
\beq
\left[i\gamma_{\mu}\partial^{\mu}-M_n^*
- g_\omega \omega \gamma^0
-g_\rho \rho \tau_3\gamma^0\right]\psi
= 0,
\eeq
\beq
 m_\sigma^2 \sigma+g_3 \sigma^3 =
-\frac {\partial M_n^*}{\partial \sigma}
\langle \bar \psi \psi \rangle,
\eeq
\beq
 m_\omega^2 \omega+c_3 \omega^3 =
g_\omega \langle \bar \psi \gamma^0 \psi \rangle,
\eeq
\beq
 m_\rho^2 \rho =
g_\rho   \langle \bar \psi \tau_3\gamma^0 \psi \rangle.
\eeq
Here, the bracket $\langle$   $\rangle$ means the expectation value
of the operator between the nuclear ground state.
For the $\omega$ and $\rho$ parts we have replaced the quark-meson
couplings $g_i^q$ by the nucleon-meson couplings $g_i$ as
$g_\omega=3g_\omega^q$ and $g_\rho=g_\rho^q$.
The effective mass $M_n^*$ and its derivative with respect to
the $\sigma$ mean field, $\partial M_n^*/\partial \sigma$,
are not trivial functions of the $\sigma$ mean field,
because $M_n^*$ depends on $\delta m_q=-g^q_\sigma\sigma$
non-trivially as shown in Fig.1.
Comparing with the RMF theory, the $\partial M_n^*/\partial \sigma$
in the QMF model is equal to the $g_\sigma$ in the RMF model.
We show in Fig.2 the quantities $g_\sigma(\sigma)/g_\sigma(0)$
as a function of $\delta m_q$.
In the case of the scalar-vector potential, the dependence of 
$g_\sigma(\sigma)$ on $\delta m_q$ is very small. 
On the other hand, in the case of the scalar potential,
$g_\sigma(\sigma)$ increases rapidly with  $\delta m_q$.

In the present model, there are five free parameters,
$g_\sigma^q$, $g_\omega^q$, $g_3$, $c_3$, and $g_\rho$, which
need to be determined.
We would like to follow the method in Ref.~\cite{MQMC}
to determine these parameters. We determine these five parameters by
reproducing five equilibrium properties of nuclear matter~\cite{TM1,MQMC}.
The equilibrium properties used here are listed in Table 1.
As for the other parameters, we take $m_\omega=783 MeV$ and
$m_\rho=770 MeV$.
We note that the variation of $m_\sigma$ at
fixed $g^q_\sigma/m_\sigma$ and $g_3/m_\sigma^4$ has no
effect on the nuclear matter properties, but
the $\sigma$ meson mass determines the range of the attractive
interaction and  affects the nuclear surface slope
and its thickness for finite nuclei
and hence the finite nuclear properties.
The mass of the $\sigma$ meson is chosen
to reproduce the charge radius of $^{40}Ca$ to be around $3.45 fm$.
The parameter sets for the four cases used in the present model 
are given in Table 2.

Using the parameters given in table 2, the nuclear matter properties
can be obtained by solving the above Euler-Lagrange equations
self-consistently. We plot in Fig.3 the energy per nucleon, $E/A$,
as functions of the nuclear matter density $\rho$.
In Fig.4, we plot the scalar potential, $U_S$, and the vector potentials,
$U_V$, as functions of density $\rho$.
The results in the RMF(TM1) model are also shown for comparison.
Definitely the results at $\rho=\rho_0=0.145 fm^{-3}$
do not change due to the construction of the parameter sets, 
but the results at different densities depend somewhat on the parameter set.

Since the QMF model uses the constituent quark model to describe
the nucleon in medium,
the nucleon properties change according to the strengths
of the mean fields.
We plot the ratio of the nucleon rms radius in medium to that in free
space as a function of density in Fig.5.
All of the four curves show  the increase of  the nucleon radius
in medium. The nucleon radius increases by about $5\% \sim 9\%$
at the normal matter density for different confinement parameters 
and the type of the potential.
It would be very interesting to calculate the EMC effect with the
change of the nucleon properties, in particular, with the increase of
the nucleon radius. If we make a comparison with the QMC 
model of Thomas and his collaborators~\cite{QMC}, a significantly swollen
nucleon radius is predicted in the present QMF model
without adding extra parameters, while almost no swelling is 
found in the orginal QMC model when the bag constant 
does not depend on density~\cite{QMC}. We note here that there are
a few modified versions of the QMC model, which could reproduce the
swollen nucleon radius by changing the bag constant~\cite{MQMC}.

\section{Properties of finite nuclei}

To study the properties of finite nuclei, the electromagnetic field must
also be included, which does not appear in infinite nuclear matter.
The construction of the nucleon inside nuclei will be rather complicated
if the variation of the meson mean fields over the nucleon volume is
considered.
We have to take some suitably averaged form for the meson mean fields
in order to make the numerical solution feasible.
Here, we use the local density approximation, which replace the meson mean
fields at the quark level by their value at the
center of the nucleon and neglect the spatial variation of the mean fields
over the small nucleon volume.
If we restrict our consideration to spherically symmetric nuclei,
the Euler-Lagrange equations are then written as
\beq
\left[i\gamma_{\mu}\partial^{\mu}-M_n^*
- g_\omega \omega(r) \gamma^0
-g_\rho \rho(r) \tau_3\gamma^0\
-e\frac{(1-\tau_3)}{2} A(r)\gamma^0
 \right]\psi
= 0,
\eeq
\beq
 \Delta\sigma(r)-m_\sigma^2 \sigma(r)-g_3 \sigma^3(r)
=\frac {\partial M_n^*}{\partial \sigma}
\langle \bar \psi \psi \rangle,
\eeq
\beq
\Delta\omega(r)-m_\omega^2 \omega(r)-c_3 \omega^3(r) =
-g_\omega \langle \bar \psi \gamma^0 \psi \rangle,
\eeq
\beq
\Delta\rho(r)-m_\rho^2 \rho(r) =
-g_\rho   \langle \bar \psi \tau_3\gamma^0 \psi \rangle,
\eeq
\beq
\Delta A(r) =
-e \langle \bar \psi \frac{(1-\tau_3)}{2}\gamma^0 \psi \rangle.
\eeq
Here, the meson mean fields are functions of $r$, which is the radial
coordinate of the nucleon center. The meson mean fields are approximated to be
constants over the small nucleon volume.
We solve the above equations self-consistently with the amount of
$M_n^*$ and $\partial M_n^*/\partial \sigma$ obtained in
the first step.

We take the same coupling constants and meson masses as used in nuclear
matter to study the properties of finite nuclei.
We follow some prescriptions in Ref.~\cite{TM1} to deal with the
center-of-mass corrections and the pair corrections.
The charge densities and the
rms charge radii are calculated by convoluting the point nucleon density
with an empirical nucleon form factor as in Ref~\cite{TM1}.
The calculated results for the binding energies per nucleon $E/A$
and the rms charge radii $R_c$ are compared with experimental values
in Table 3.
For the case of the scalar-vector potential, the calculated results
for $E/A$ and $R_c$ are almost same as the RMF(TM1) results.
This is because the derivative $\partial M_n^*/\partial \sigma$
is almost unchanged with the $\sigma$ mean field as shown in Fig.2.
On the other hand, for the case of the scalar potential, 
$E/A$ are somewhat underestimated.
This is caused by the increase of $g_\sigma$
with the $\sigma$ mean field as shown in Fig.2.
We need, instead, the decrease of the effective $g_\sigma$
in order to achieve good results.
In Table 4, the calculated spin-orbit splittings for $^{40}Ca$
and $^{208}Pb$ are presented.
The results are similar to those of the RMF(TM1).
This is due to the fact that the effective mass was used to obtain
the coupling constants.
In Fig.6 we show calculated charge density distributions for $^{40}Ca$
in the QMF model and compare with the experimental distribution~\cite{EXP}
and the results in the RMF(TM1) model.
Even though there are still some discrepancies between the results in the
QMF model and the experimental values,
we consider the QMF model provides reasonable results for finite nuclei.

\section{Conclusion}

We have developed the quark mean field (QMF) model to describe the
change of nucleon properties in nuclei and at the same time the
properties of nuclear matter and finite nuclei.
We have used the constituent quark model for the nucleon, which
naturally allows the direct coupling of $\sigma$, $\omega$, and
$\rho$ mesons with quarks.
The mean field Lagrangian at the nucleon level reflects the direct
coupling of mesons with quarks merely through the appearance of
the effective nucleon mass, which is a function of the $\sigma$ mean field.
With our model setting and our  parameter choices,
we can perform the numerical calculations for nuclear matter
and finite nuclei.

We have investigated the QMF model with different types of the confinement
potential. We have taken the mean field Lagrangian with nonlinear terms for
both $\sigma$ and $\omega$ mesons.
The comparison between the QMF model and the RMF(TM1) model has been done.
The QMF model can provides a significantly swollen nucleon radius
in nuclear medium.
At the normal matter density, the nucleon radius increases by about
$5\% \sim 9\%$.
We calculate also the properties of finite nuclei, the resulting
binding energies per nucleon $E/A$ and charge radii $R_c$
are close to the experimental values, as well as the spin-orbit
splittings in the QMF model are also satisfactory.

\section*{Acknowledgments}
H. Shen would like to thank A. Hosaka for fruitful discussions
and helpful suggestions.
This work was supported in part by National Natural Science
Foundation of China.


\newpage
\section*{Figure captions}

\begin{description}

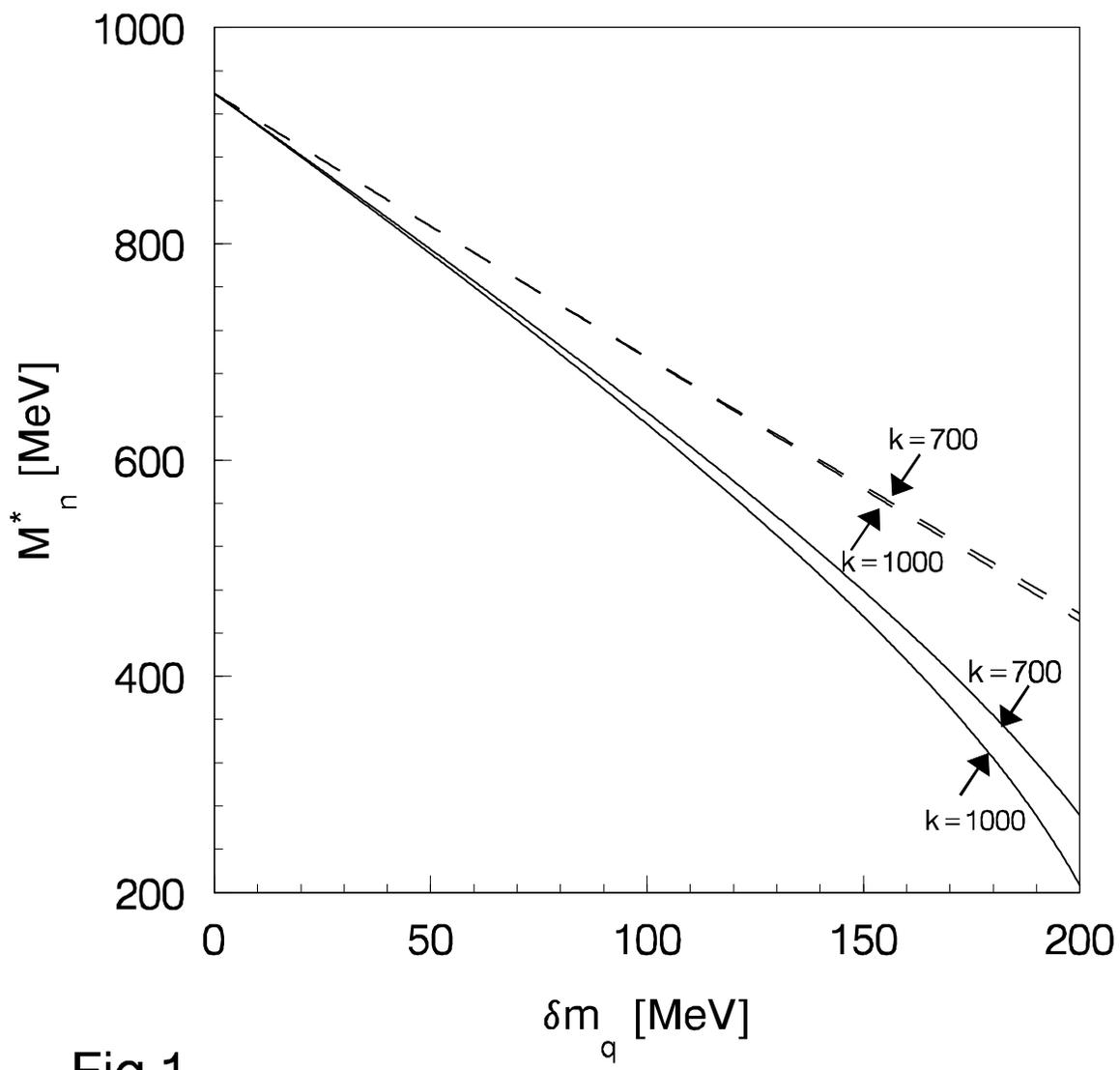
 \item[Figure 1:]  The effective nucleon mass $M_n^*$ as
    functions of the quark mass correction $\delta m_q$.
    The results in the QMF model with
    $\chi_c=\frac{1}{2}kr^2$ are shown by solid curves,
    while those with $\chi_c=\frac{1}{2}kr^2 (1+\gamma^0)/2$
    are shown by dashed curves.
    For each potential shown are the two results for two confining strengths.
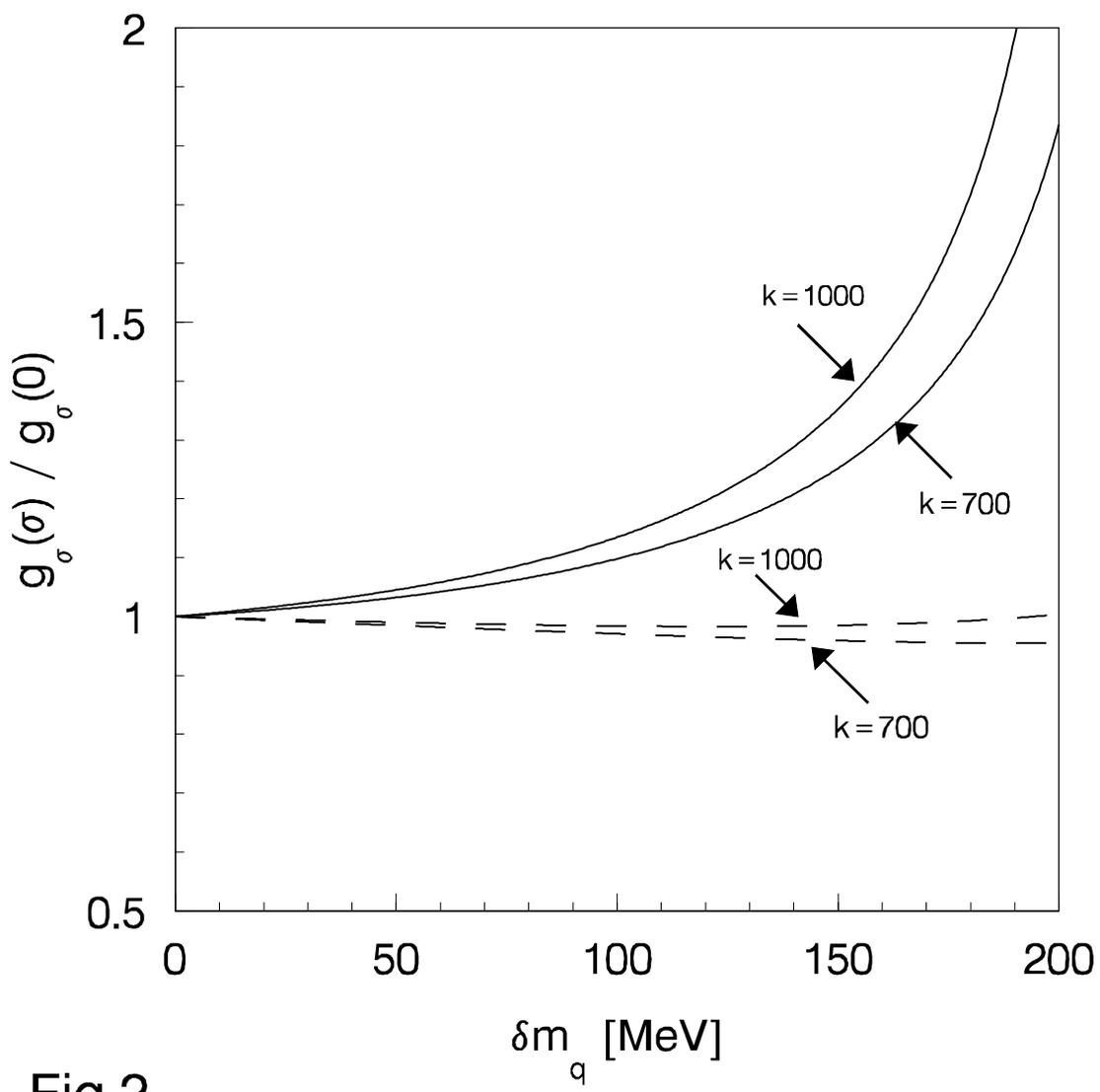
 \item[Figure 2:]  The ratios of the $\sigma$-nucleon coupling in medium,
    $g_\sigma (\sigma)=\partial M_n^*/\partial \sigma$,
    to that in free space,  $g_\sigma (0)$,
    as functions of the quark mass correction $\delta m_q$.
    $\delta m_q$ is connected with the $\sigma$ mean field as
    $\delta m_q=-g_{\sigma}^q\sigma$.
    The curves are labeled as in Fig.1.
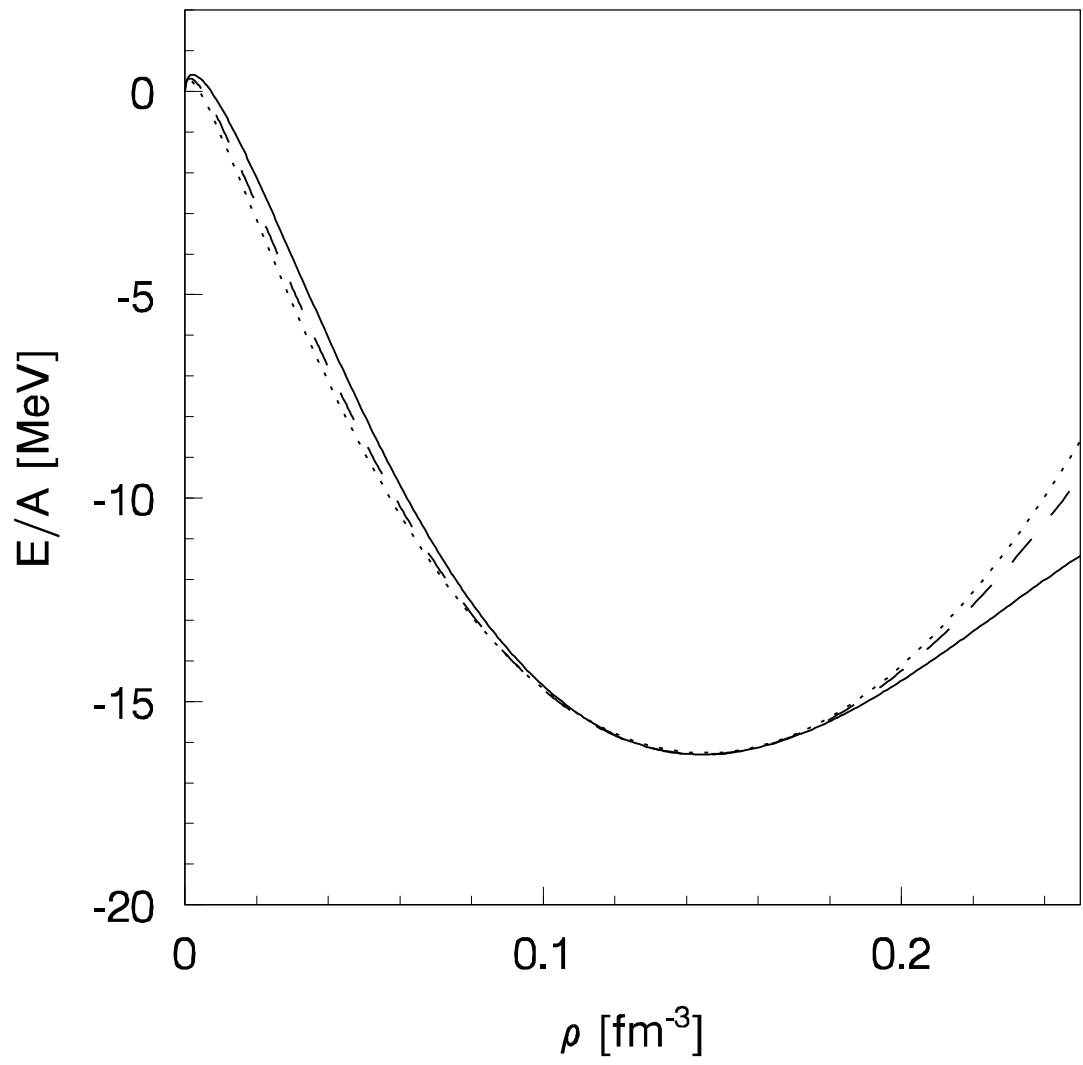
 \item[Figure 3:]  The energy per nucleon, $E/A$, as functions of the
    nuclear matter density $\rho$.
    The results in the QMF model with
    $\chi_c=\frac{1}{2}kr^2$ are shown by solid curves,
    while those with $\chi_c=\frac{1}{2}kr^2 (1+\gamma^0)/2$
    are shown by dashed curves with $k=700$.
    The results in the RMF(TM1) model are plotted by dotted curves
    for comparison.
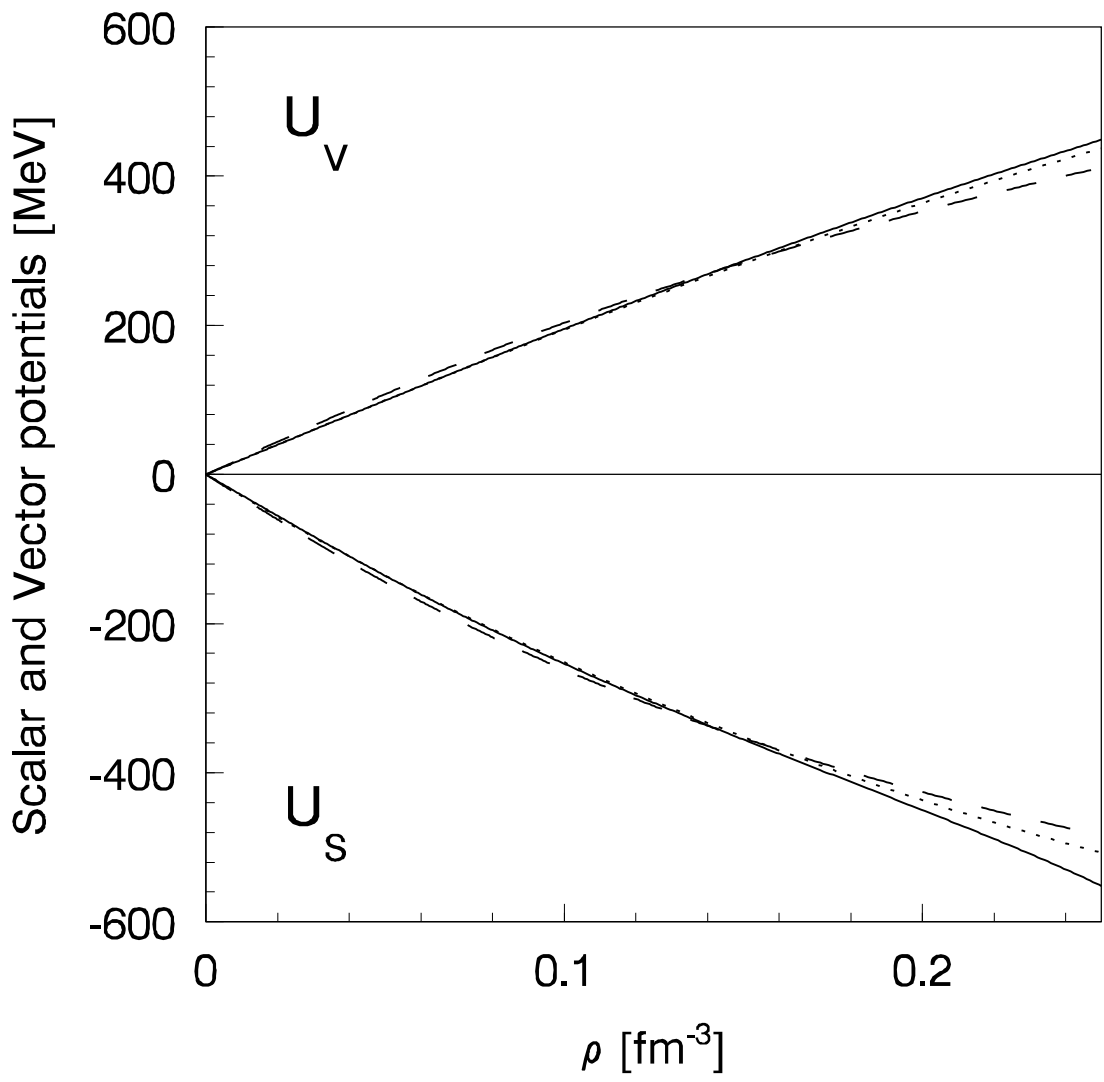
 \item[Figure 4:]  The scalar potential, $U_S$, and the vector
    potential, $U_V$, as functions of
    the nuclear matter density $\rho$.
    The curves are labeled as in Fig.3.
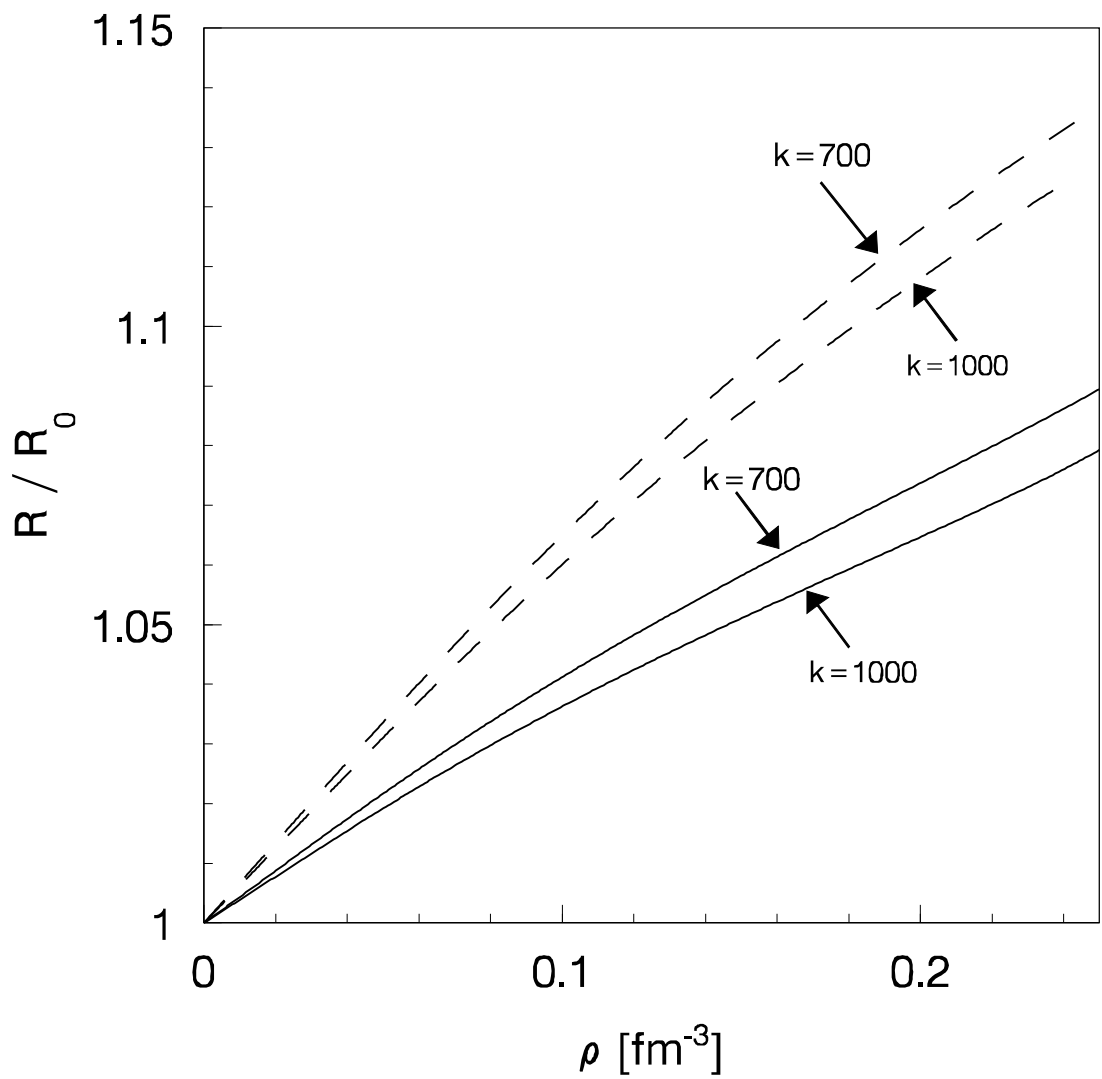
 \item[Figure 5:]  The ratios of the nucleon rms radius $R$
    to that in free space $R_0$ as functions of the nuclear
    matter density $\rho$.
    The curves are labeled as in Fig.1.
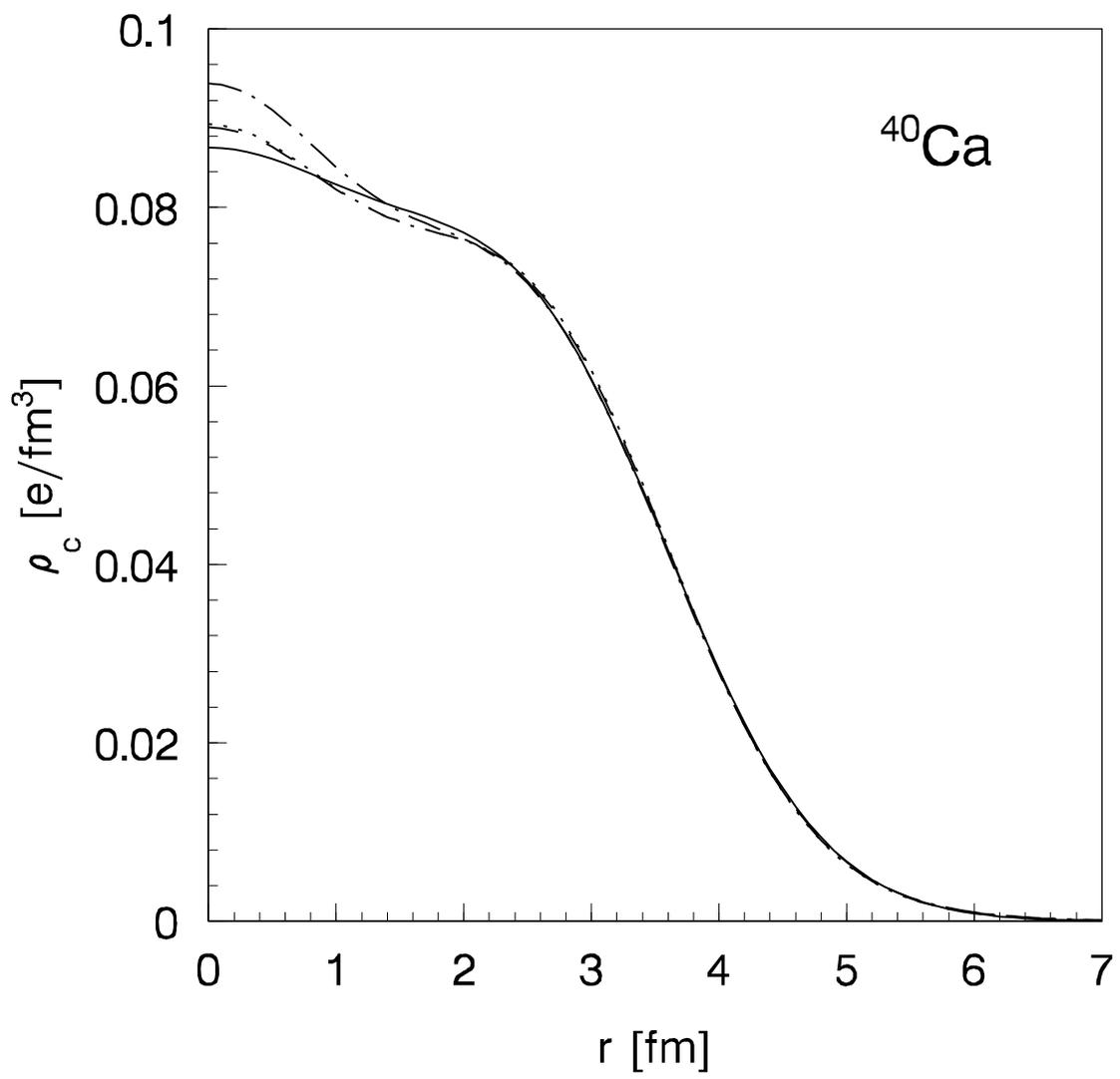
 \item[Figure 6:]  The charge density distributions for $^{40}Ca$
    compared with the experimental data (solid curve)~\cite{EXP}.
    The dash-dotted and dashed curves are the results in the QMF model
    with $\chi_c=\frac{1}{2}kr^2$ $(k=700)$ and
         $\chi_c=\frac{1}{2}kr^2 (1+\gamma^0)/2$ $(k=700)$, respectively.
    The results in the RMF(TM1) model are plotted by dotted curves
    for comparison.
\end{description}

\newpage
\begin{table}
    \caption{The nuclear matter properties used to determine the
             five free parameters in the present model.
             The saturation density and the energy per particle are denoted by
             $\rho_0$ and $E/A$, and the incompressibility by $k$,
             the effective mass by $M_n^*$ and the symmetry energy
             by $a_{sym}$.}
\vspace{0.5cm}
\begin{center}
\begin{tabular}{ccccc} \hline
\hline
  $\rho_0$  & $E/A$  & $k$   & $M_n^*/M_n$ & $a_{sym}$   \vspace{-0.3cm}\\
$(fm^{-3})$ &$(MeV)$ &$(MeV)$&             & $(MeV)$     \\  \hline
    0.145   & -16.3  & 280   & 0.63        & 35          \\    \hline
\hline
\end{tabular}
\end{center}
\end{table}
\vspace{1cm}

\begin{table}
    \caption{The parameters in the QMF model are listed. For comparison,
             the parameters in the RMF(TM1) model are also presented.}
\vspace{0.5cm}
\begin{center}
\begin{tabular}{ll|ccccccc} \hline
\hline
   Model &  & $g^q_\sigma$ & $g^q_\omega$  & $g_2$ & $g_3$ & $c_3$
   &  $g_\rho$ & $m_\sigma$ \vspace{-0.3cm}    \\
         &  &              &               & $(fm^{-1})$&  &
   &           & $(MeV)$                       \\  \hline
   QMF   & $k=700$
  & 3.14 & 4.20 & 0 & 50.7 & 53.6 & 4.3 & 470  \\ \cline{2-9}
   $\chi_c=\frac{1}{2}kr^2$ & $k=1000$
  & 2.98 & 4.17 & 0 & 52.8 & 36.4 & 4.3 & 460  \\ \hline
   QMF   & $k=700$
  & 4.18 & 4.42 & 0 & 36.8 &214 & 4.3 & 515  \\ \cline{2-9}
   $\chi_c=\frac{1}{2}kr^2 (1+\gamma^0)/2$ & $k=1000$
  & 4.11 & 4.38 & 0 & 36.4 & 167 & 4.3 & 510  \\ \hline
   RMF (TM1) &
  & $g_\sigma=10.029$ & $g_\omega=12.614$
  & -7.2325 & 0.6183 & 71.308 & 4.6322 & 511.2 \\  \hline
\hline
\end{tabular}
\end{center}
\end{table}
\vspace{1.0cm}

\begin{table}
    \caption{The binding energies per nucleon $E/A$ and the rms charge
             radii $R_c$ in the present model compared with the
             results in the RMF(TM1) model and the experimental values~\cite{TM1}.}
\vspace{0.5cm}
\begin{center}
\begin{tabular}{ll|cccc|cccc} \hline
\hline
   Model &  &  & $E/A$ & $(MeV)$ &  &  & $R_c$ & $(fm)$  &  \\
   \cline{3-10}
         &  & $^{40}Ca$ & $^{48}Ca$ &  $^{90}Zr$ &  $^{208}Pb$
            & $^{40}Ca$ & $^{48}Ca$ &  $^{90}Zr$ &  $^{208}Pb$  \\  \hline
   QMF   & $k=700$
  & 7.53 & 7.66 & 7.92 & 7.36 & 3.45 & 3.46 & 4.28 & 5.53  \\ \cline{2-10}
   $\chi_c=\frac{1}{2}kr^2$ & $k=1000$
  & 7.32 & 7.46 & 7.75 & 7.24 & 3.45 & 3.46 & 4.28 & 5.53  \\ \hline
   QMF   & $k=700$
  & 8.35 & 8.43 & 8.54 & 7.81 & 3.44 & 3.46 & 4.28 & 5.54  \\ \cline{2-10}
   $\chi_c=\frac{1}{2}kr^2 (1+\gamma^0)/2$ & $k=1000$
  & 8.21 & 8.30 & 8.43 & 7.73 & 3.44 & 3.46 & 4.27 & 5.53  \\ \hline
   RMF (TM1) &
  & 8.62 & 8.65 & 8.71 & 7.87 & 3.44 & 3.45 & 4.27 & 5.53  \\ \hline
   Exp.      &
  & 8.55 & 8.67 & 8.71 & 7.87 & 3.45 & 3.45 & 4.26 & 5.50  \\ \hline
\hline
\end{tabular}
\end{center}
\end{table}
\vspace{1cm}

\begin{table}
    \caption{The spin-orbit splittings for $^{40}Ca$ and $^{208}Pb$
             in the present model compared with the results
             in the RMF(TM1) model and the experimental values~\cite{EXPSO}.
             All quantities are in $MeV$.}
\vspace{0.5cm}
\begin{center}
\begin{tabular}{ll|cc|cc} \hline
\hline
   Model &  & $^{40}Ca$ &  & $^{208}Pb$ &  \\
   \cline{3-6}
         &  &     Proton &   Neutron &    Proton &    Neutron \\
            &
            & $(1d_{5/2}-1d_{3/2})$
            & $(1d_{5/2}-1d_{3/2})$
            & $(1g_{9/2}-1g_{7/2})$
            & $(2f_{7/2}-2f_{5/2})$
\\   \hline
   QMF   & $k=700$
  & -5.8 & -5.9 & -3.5 & -1.8  \\ \cline{2-6}
   $\chi_c=\frac{1}{2}kr^2$ & $k=1000$
  & -5.8 & -5.8 & -3.5 & -1.8  \\ \hline
   QMF   & $k=700$
  & -5.6 & -5.6 & -3.3 & -1.9  \\ \cline{2-6}
   $\chi_c=\frac{1}{2}kr^2 (1+\gamma^0)/2$ & $k=1000$
  & -5.7 & -5.8 & -3.4 & -1.9 \\ \hline
   RMF (TM1) &
  & -5.7 & -5.7 & -3.4 & -1.8 \\ \hline
   Exp.      &
  & -7.2 & -6.3 & -4.0 & -1.8  \\ \hline
\hline
\end{tabular}
\end{center}
\end{table}

\end{document}